\documentclass[11pt]{article}
\usepackage{graphicx}
\usepackage{authblk}

\usepackage[shortlabels]{enumitem}
\usepackage{amsmath} 
\usepackage{amssymb} 
\usepackage{mathrsfs} 
\usepackage{mathtools} 
\usepackage{amsthm}
\usepackage{bm}
\usepackage{thmtools}
\usepackage{thm-restate}
\usepackage{appendix}
\usepackage[in]{fullpage}
\usepackage{caption}
\usepackage{subcaption}
\usepackage{graphicx}
\usepackage[table,xcdraw]{xcolor}
\usepackage{wrapfig}
\usepackage{floatrow}
\usepackage{multirow}
\usepackage{makecell}
\usepackage{nicefrac}
\usepackage{dsfont}

\usepackage[ruled,vlined,linesnumbered]{algorithm2e}
\definecolor{forestgreen}{rgb}{0.13, 0.55, 0.13}

\SetCommentSty{mycommfont}
\SetKwInOut{Input}{input}
\SetKwInOut{Output}{output}
\DontPrintSemicolon

\usepackage[backend=biber,style=alphabetic,maxbibnames=9,maxcitenames=5]{biblatex}
\definecolor{ForestGreen}{rgb}{0.0333,0.4451,0.0333}
\definecolor{DarkRed}{rgb}{0.65,0,0}
\definecolor{Red}{rgb}{1,0,0}
\usepackage[linktocpage=true,
colorlinks,
linkcolor=DarkRed,citecolor=ForestGreen,
bookmarks,bookmarksopen,bookmarksnumbered]{hyperref}

\usepackage{cleveref}
\usepackage{thm-restate}
\usepackage{tcolorbox}
\usepackage[normalem]{ulem} 

\newtheorem{theorem}{Theorem}[section]
\newtheorem{lemma}[theorem]{Lemma}
\newtheorem{problem}[theorem]{Problem}

\newtheorem{claim}[theorem]{Claim}

\newtheorem{proposition}[theorem]{Proposition}

\newtheorem{definition}[theorem]{Definition}

\DeclarePairedDelimiter\abs{\lvert}{\rvert}

\newcommand*\ceil[1]{\left\lceil #1 \right\rceil}

\newcommand{\OPT}{\text{OPT}}

\newcommand{\fami}{\mathcal{F}}
\newcommand{\R}{\mathbb{R}}

\newcommand{\ralg}{\mathcal{A}}

\newif\ifcomments
\commentstrue

\ifcomments
\usepackage[colorinlistoftodos,prependcaption,textsize=tiny]{todonotes}

\definecolor{dnotecol}{rgb}{0.20, 0.50, 0.80}
\newcommand{\dnote}[1]{{\textcolor{dnotecol}{\textbf{DH:} #1}}}
\newcommand{\dnotein}[1]{\todo[linecolor=red,backgroundcolor=dnotecol!25,bordercolor=red,inline]{\textbf{DH:~}#1}}

\definecolor{imAboutIt}{rgb}{1, 0.5, .31}

\definecolor{wnotecol}{rgb}{0.90, 0.00, 0.00}
\newcommand{\wnote}[1]{{\textcolor{wnotecol}{\textbf{WU:} #1}}}

\else 

\newcommand{\dnote}[1]{}
\newcommand{\dnotein}[1]{}
\newcommand{\dqtodo}[1]{}
\newcommand{\dqtodoin}[1]{}
\newcommand{\wnote}[1]{}

\fi 

\title{Weighted and Unweighted Online Connectivity Augmentation \\ via Geometric Hitting Set}

\author{
    Daniel Hathcock\thanks{GE Vernova, Advanced Research Center, Niskayuna, NY, USA}
    \qquad
    Aditya Subramanian\thanks{University of Waterloo, ON, Canada}
    \qquad
    Seeun William Umboh\thanks{University of Melbourne, VIC, Australia}
    \qquad
    Michael Zlatin\thanks{Department of Computer Science, Pomona College, CA, USA}
}
\date{}

\begin{document}

\maketitle

\begin{abstract}
We study the Online Connectivity Augmentation Problem (OCAP) with cut arrivals. In this problem, we are given a $k$-edge-connected graph $G = (V,E)$ with $|V| = n$ and a set of links $L \subseteq \binom{V}{2}$. In each iteration, a minimum cut arrives to be augmented, and we seek to maintain a smallest set of links crossing all such cuts. 
This problem is a special case of online hitting set in which the ground set is $L$ and a sequence of sets to be hit arrive from $\mathcal{C} = \{\delta_L(C) : |\delta_E(C)| = k\}$. As such, OCAP admits a $O(\log^2 n)$-competitive algorithm using the classical algorithm for Online Set Cover of Alon, Awerbuch, Azar, Buchbinder, and Naor.
If $k$ is odd, then OCAP reduces to the case where $G$ is a tree. This case was studied by Naor, Umboh, and Williamson, who gave an optimal $O(\log n)$-competitive algorithm, which even holds in weighted setting. Recently, Garg and Subramanian gave an $O(\log n)$-competitive algorithm for unweighted OCAP which holds in the popular node-arrival model, but not for cut-arrivals. 
No better than $O(\log^2 n)$-competitive algorithm is known for weighted OCAP. In this paper, we provide a nearly optimal $O(\log n \log \log n)$-competitive algorithm for weighted OCAP under cut-arrivals and an optimal $O(\log n)$-competitive algorithm for unweighted OCAP with cut-arrivals.
We also give an $O(1)$ algorithm for Online Cycle Augmentation under node-arrivals. Since no $o(\log n)$-competitive algorithm is possible on a cycle with cut-arrivals, this establishes a separation between the best competitive ratio achievable in the two arrival models.
\end{abstract}

\section{Introduction}
The defining goal of the field of resilient network design is to construct a network which is robust to failures of its elements. Often, the network designer does not have the luxury of designing their network from scratch, rather there is an existing legacy network in place which we seek to augment by adding additional edges, called links, in the most efficient way. These problems are called Network Augmentation Problems, and have motivated a substantial amount of work in approximation algorithms, combinatorial optimization and polyhedral theory. 

A common measure of the resiliency of a network to edge failures is its \textbf{edge-connectivity}. Given a graph $G = (V,E)$, a pair of vertices $u,v \in V$ are $k$-edge-connected if the size of any cut separating $u$ and $v$ is at least $k$. Similarly, $G$ is (globally) $k$-edge-connected if every pair of vertices is. This motivates the following problem, called the Weighted Connectivity Augmentation Problem, or WCAP. 

\begin{problem}[WCAP]
Given a $k$-edge-connected graph $G=(V,E)$ and a set of links $L \subseteq \binom{V}{2}$ with nonnegative costs $c:L\to\mathbb{R}_{\geq 0}$, find a minimum-cost subset $F\subseteq L$ such that $(V,E\cup F)$ is $(k+1)$-edge-connected.
\end{problem}

It is useful to observe that this problem can be formulated as a special case of hitting set, where a feasible solution corresponds exactly to a set of links $F \subseteq L$ hitting all $k$-cuts of $G$. Here, a link $\ell$ \textbf{hits }a cut $C \subseteq V$ if it crosses that cut, i.e.~if $\ell \in \delta_L(C)$. This observation allows us to replace the graph $G$ by another, as long as the structure of its minimum cuts remains the same. In particular, Dinitz, Karzonov, and Lomonosov~\cite{DinitzKarzanovLomonosov1976} showed that for any $k$-edge-connected graph $G$, there exists an (efficiently computable) cactus graph $H$ for which there is a one-to-one correspondence between their minimum cuts. Furthermore, if $k$ is odd, then $H$ can be chosen to be a tree. Thus, the WCAP problem reduces to the case where $G$ is a cactus (for all $k$) and where $G$ is a tree (if $k$ is odd) these problems are called WCacAP and WTAP respectively.

There is much work on the offline variants of CAP and TAP. However, there has also been a strong interest in Connectivity Augmentation Problems in other models of computation, including the online and streaming settings \cite{GuptaKrishnaswamyRavi2012, NaorUW22, aditya, jin_et_al:LIPIcs.ICALP.2024.93, chekuri_et_al:LIPIcs.APPROX/RANDOM.2025.4}. We focus on the Online Weighted Connectivity Augmentation Problem (OWCAP) with cut-arrivals, defined as follows.

\begin{problem}[OWCAP; cut-arrivals]
Given a $k$-edge-connected graph $G=(V,E)$, and a set of links $L \subseteq \binom{V}{2}$ on $V$ with non-negative costs $c:L\to\mathbb{R}_{\geq 0}$, a sequence of $k$-cuts of $G$ arrive online. Upon each arrival, we must irrevocably purchase links so that every $k$-cut that has arrived so far is crossed by a link we have purchased. The objective is to minimize the total cost of purchased links.
\end{problem}

There is also the problem of Online Connectivity Augmentation with node-arrivals in which pairs of vertices $(u_t,v_t) \in \binom{V}{2}$, referred to as \emph{terminals}, arrive online. Upon each arrival, we must irrevocably purchase links so that every requested pair is $(k+1)$-edge-connected in the augmented graph. Notice that each terminal-pair request corresponds to covering a collection of cut requests---namely those $k$-cuts which separate $u$ from $v$ in $G$. Hence, any algorithm for OCAP in the cut-arrival model can be used to solve OCAP with node-arrivals by processing all $k$-cuts that separate pairs of requested vertices. However, the converse is not true: an algorithm which handles node-arrivals may not apply to the cut-arrival setting. 

As a special case of Online Hitting Set, the OWCAP with cut-arrivals problem readily admits an $O(\log^2 n)$-competitive algorithm by applying the classical algorithm of \cite{AlonAABN09}. In particular, \cite{AlonAABN09} provides an algorithm for weighted online hitting set which is $O(\log m\log N)$-competitive, where $m$ is the number of sets to be hit and $N$ is the size of the ground set of elements. In OCAP with cut arrivals, our ground set is $L$ and the sets to be hit arrive from $\mathcal{C} = \{\delta_L(C) : |\delta_E(C)| = k\}$, each set corresponding to a min-cut of $G$. It follows from \cite{DinitzKarzanovLomonosov1976} that the number of minimum cuts in any graph with $n$ vertices is at most $n^2$, hence, we have $N = |L| \leq n^2$ and $m \leq n^2$ yielding an $O(\log^2 n)$ competitive ratio overall. 

Following this, subsequent work has obtained improved algorithms in various important special cases. For example, Naor, Umboh and Williamson~\cite{NaorUW22} considered the special case of Online WTAP, in which $G$ is a tree. They essentially resolve this case, giving a deterministic $O(\log n)$-competitive algorithm for Online WTAP, matching a lower bound showing that no $o(\log n)$-competitive ratio can be achieved~\cite{ GuptaKrishnaswamyRavi2012}. More recently, Garg and Subramanian~\cite{aditya} considered the general case of Online CAP, giving an optimal $O(\log n)$-competitive algorithm for unweighted CAP in the node-arrival model. 
However, their techniques do not extend to cut arrivals. Interestingly, as observed in~\cite{NaorUW22}, the node-arrival model and the cut-arrival model are equivalent for Online WTAP.
No better than $O(\log^2n)$ is known for weighted OCAP in either arrival model. Thus, our work is motivated by two central questions:

\begin{enumerate}[label={}]
    \item 
    \textbf{Question 1:} Is there a separation between the best achievable
    competitive ratios in the node-arrival and cut-arrival
    models of Online Connectivity Augmentation?

    \item
    \textbf{Question 2:} Can we obtain an optimal $O(\log n)$-competitive
    algorithm for weighted Online Connectivity Augmentation?
\end{enumerate}

More generally, for which special cases of Online Hitting Set
can we exploit the underlying structure to obtain improved
competitive ratios? In the next section, we describe how
we address each of these questions.

\subsection{Our Results}

We essentially resolve Question 1 and resolve Question 2 up to a factor of $\log \log n$. Our first result is a nearly optimal competitive ratio for OWCAP with cut arrivals.

\begin{restatable}{theorem}{thmOWCAP}\label{thm:OWCAP}
    There is a randomized $O(\log n \log \log n)$-competitive algorithm for OWCAP with cut arrivals.
\end{restatable}

Since the cut arrivals model subsumes the node arrival model, this also implies a $O(\log n\log\log n)$-competitive result for node arrivals, improving the earlier $O(\log^2 n)$ bound.

We also obtain an optimal $O(\log n)$-competitive algorithm for unweighted OCAP with cut arrivals. This generalizes the recent result of \textcite{aditya} who achieved a matching result only for node arrivals. Furthermore, our approach is distinct, and hence also provides an alternate algorithm for the node arrival case, with matching guarantees.

\begin{theorem}\label{thm:OCAP}
    There is an $O(\log n)$-competitive algorithm for unweighted OCAP under cut arrivals.
\end{theorem}

Thus, \Cref{thm:OCAP} gives a partial answer to Question 1, namely that there is no difference in complexity between node-arrival and cut-arrival model in the unweighted case of OCAP for general graphs. However, we show that for unweighted OCAP where $G$ is a cycle, which we call OCycAP, there is a separation between the two models. 
 
\begin{theorem}\label{thm:NOCAP}
    There is a $O(1)$-competitive algorithm for unweighted OCycAP under node-arrivals. However, there is no $o(\log n)$-competitive algorithm for unweighted OCycAP under cut-arrivals, even fractionally.
\end{theorem}

Along the way to proving \Cref{thm:OWCAP} and \Cref{thm:OCAP} we establish a surprising connection between connectivity augmentation problems and a particular case of 2D geometric hitting set, which we call Online TL-Rectangle hitting set. We elaborate on this further in the later sections, but we state the result here, as it is interesting in its own right.
   
\begin{theorem}\label{thm:introTLrect}
    There is a randomized $O(\log |P| \log \log |P|)$ competitive algorithm for online hitting set with $P$ points where each arriving shape is the union of a leftless rectangle and a topless rectangle.
\end{theorem}

As far as the authors can tell, \Cref{thm:cycletoTL} is the only result for any non-trivial special case of general weighted geometric online hitting set which beats the doubly logarithmic factor obtainable via~\cite{AlonAABN09} for general weighted hitting set.

\subsection{Our Techniques}
\label{sec:techniques}
Our approach to Online Connectivity Augmentation with cut-arrivals consists of three phases: reduction to a cycle, then reduction to a particular case of geometric hitting set we call TL-Rectangle Hitting Set, then solving TL-Rectangle Hitting Set with the appropriate bounds.

\subsubsection{From arbitrary graphs to cycles}
First we reduce the general OCAP problem (both weighted and unweighted) to the same problem on a cycle. This reduction is a standard tool when dealing with Connectivity Augmentation Problems in the weighted case. Essentially, one can introduce links of weight 0 to ``unfold" any cactus into a cycle, where the minimum cuts in the cycle which are not crossed by 0 cost links are in one-to-one correspondence with the minimum cuts of the original cactus (c.f.~Theorem 4 in~\cite{GGAS21}, Lemma 2.2 in~\cite{TraubZenklusen2023WCAP}). 

Since this introduces links of cost 0, the reduction does not go through for unweighted CAP. However, we show that in the online setting with cut arrivals, we may always assume that the graph to be augmented is a cycle. 

\begin{theorem}\label{thm:redtocycle}
    If there is an $\alpha$-competitive algorithm for OCycAP with cut-arrivals, then there is an $\alpha$-competitive algorithm for OCAP with cut-arrivals. Similarly, if there is an $\alpha$-competitive algorithm for OWCycAP with cut-arrivals, then there is an $\alpha$-competitive algorithm for OWCAP with cut arrivals.
\end{theorem}

To prove \Cref{thm:redtocycle}, we utilize the Theorem on Circumference Disposition of Dinitz, Karzonov, and Lomonosov~\cite{DinitzKarzanovLomonosov1976}. This result states that the vertices of any graph can be arranged in a cyclic order so that any minimum cut in $G$ splits the vertices into two arcs of this cycle. In other words, for any graph $G=(V,E)$ there exists a cycle $H=(V,E')$ such that if $S \subseteq V$ is a min-cut in $G$, then $S$ is a min-cut $H$. While there may be additional min-cuts in $H$ which do not correspond to min-cuts in $G$, in the cut-arrival model, these cuts simply never request to be augmented.

We emphasize that \Cref{thm:redtocycle} \textit{only }applies to the cut-arrival model. It is not true for node-arrivals, as \Cref{thm:NOCAP} demonstrates. In particular, there is a constant competitive algorithm for unweighted OCycAP in the node-arrival model, but no $o(\log n)$-competitive algorithm exists for unweighted OCAP in the node-arrival model. 

\subsubsection{From cycle-cuts to TL-rectangles}
Our second step is to cast OCycAP as a 2D geometric hitting set problem we call TL-rectangle hitting set. This problem is an Online 2D geometric hitting set problem in which each arriving shape is the union of two particularly oriented 3-sided rectangles - one is topless and the other is leftless. These shapes are defined as follows:

\begin{definition}
    A topless rectangle is a region $R^T = \{(x,y) \in \mathbb{R}^2 : a \leq x \leq b, c \leq y\}$ for some $a,b,c \in \mathbb{R}$. Equivalently, $R^T = [a,b] \times [c, \infty)$. A leftless rectangle is a region $R^L = \{(x,y) \in \mathbb{R}^2 : x \leq a, b \leq y \leq c\} = (-\infty, a] \times [b,c]$ for some $a,b,c \in \mathbb{R}$. 
\end{definition}

Then, the Online TL-rectangle hitting set problem is the standard geometric hitting set problem in 2D where each arriving shape is the union of a topless and leftless rectangle.

\begin{problem}[Online TL-Rectangle Hitting Set]\label{prob:TL-hitting-set}
    We are given a finite point set $P \subseteq \mathbb{R}^2$. In each time step, a shape $S \subseteq \mathbb{R}^2$ arrives to be hit where $S = R^L \cup R^T$ is the union of a leftless rectangle $R^L$ and a topless rectangle $R^T$. Upon each arrival, we may irrevocably purchase points to maintain a set $F \subseteq P$ which hits all shapes which have arrived so far. The goal is to minimize $|F|$. In the weighted case, each point has an associated non-negative cost $c_p$ and the goal is to minimize the total cost of $F$. 
\end{problem}

Perhaps surprisingly, we show that OCycAP with cut-arrivals reduces to Online TL-rectangle hitting set in both the weighted and unweighted case.  

\begin{theorem}\label{thm:cycletoTL}
    If there is an $\alpha$-competitive algorithm for (weighted) Online TL-Hitting Set, then there is an $\alpha$-competitive algorithm for (weighted) OCycAP.
\end{theorem}

To see why this is true, consider any  instance of OCycAP with cycle $G = (V,E)$ and links $L$. We will represent each link as a point in $\mathbb{R}^2$, and each 2-cut of $G$ will correspond to the union of a topless and leftless rectangle. We will do this in such a way so that each link $\ell$ crosses the min-cut $S$ if and only if its associated point is contained in the shape associated to $S$.

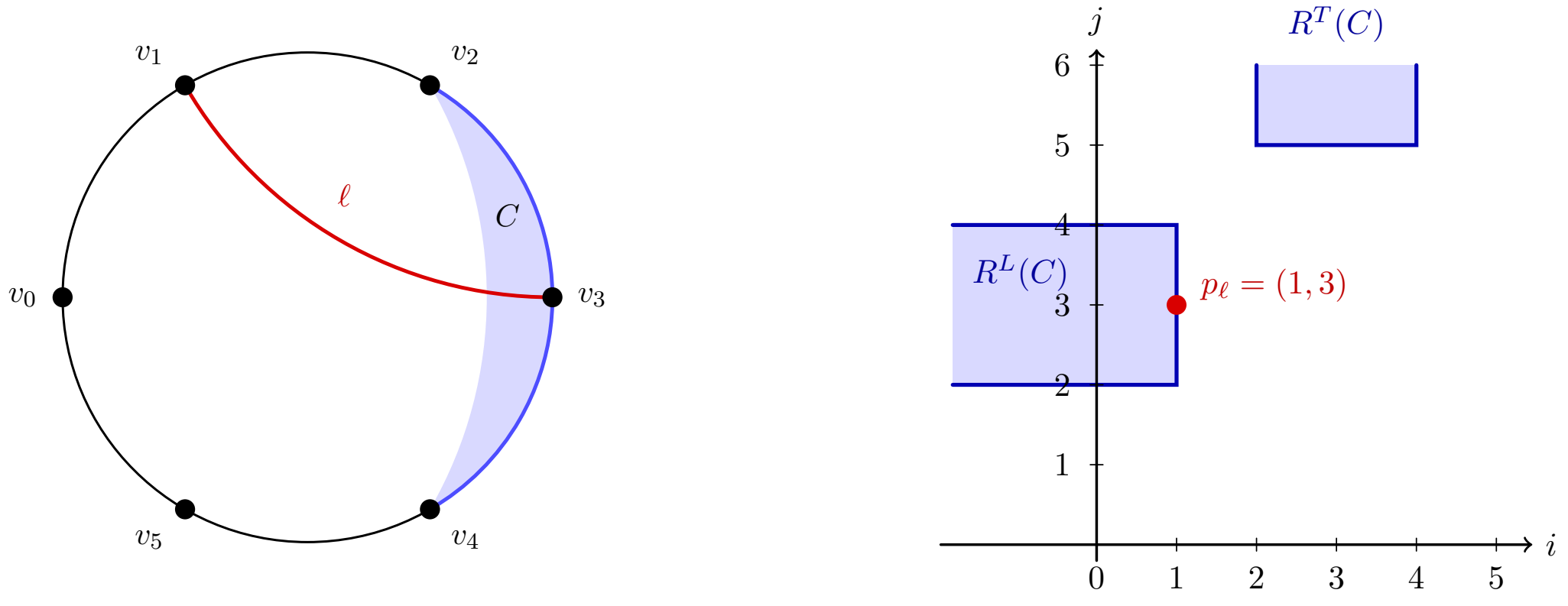
\begin{figure}[ht]
\centering
\begin{minipage}[c]{0.33\linewidth}
\centering
\resizebox{\linewidth}{!}{
\begin{tikzpicture}[line cap=round]
\coordinate (v0) at (180:3);
\coordinate (v1) at (120:3);
\coordinate (v2) at (60:3);
\coordinate (v3) at (0:3);
\coordinate (v4) at (-60:3);
\coordinate (v5) at (-120:3);
\fill[blue!15]
    (v2) arc[start angle=60, end angle=-60, radius=3]
    arc[start angle=-30, end angle=30, radius={3*sqrt(3)}]
    -- cycle;
\draw[black, thick] (0,0) circle (3);
\draw[blue!70, line width=1.3pt]
    (v2) arc[start angle=60, end angle=-60, radius=3];
\draw[red!85!black, line width=1.3pt]
    (v1) arc[start angle=-150, end angle=-90, radius={3*sqrt(3)}];
\node[red!75!black] at (0.45,1.25) {$\ell$};
\foreach \i in {0,...,5} {
    \fill[black] (v\i) circle (3.5pt);
}
\node[left=5pt]        at (v0) {$v_0$};
\node[above left=5pt]  at (v1) {$v_1$};
\node[above right=5pt] at (v2) {$v_2$};
\node[right=5pt]       at (v3) {$v_3$};
\node[below right=5pt] at (v4) {$v_4$};
\node[below left=5pt]  at (v5) {$v_5$};
\node at (2.45,1) {$C$};
\end{tikzpicture}%
}
\end{minipage}\hspace{0.04\linewidth}%
\hspace{2cm}
\begin{minipage}[c]{0.33\linewidth}
\centering
\resizebox{\linewidth}{!}{%
\begin{tikzpicture}[x=0.85cm,y=0.85cm,line cap=round,
    every node/.style={font=\small}]
\fill[blue!15] (-1.8,2) rectangle (1,4);
\draw[blue!70!black, very thick]
    (-1.8,2) -- (1,2) -- (1,4) -- (-1.8,4);
\fill[blue!15] (2,5) rectangle (4,6);
\draw[blue!70!black, very thick]
    (2,6) -- (2,5) -- (4,5) -- (4,6);
\draw[->,thick] (-1.95,0) -- (5.45,0) node[right] {$i$};
\draw[->,thick] (0,-0.2) -- (0,6.2) node[above] {$j$};
\foreach \i in {0,...,5} {
    \draw (\i,-0.08) -- (\i,0.08);
    \node[below=3pt] at (\i,0) {$\i$};
}
\foreach \j in {1,...,6} {
    \draw (-0.08,\j) -- (0.08,\j);
    \node[left=4pt] at (0,\j) {$\j$};
}
\node[blue!60!black] at (-0.95,3.4) {$R^L(C)$};
\node[blue!60!black,anchor=south] at (3,6.13) {$R^T(C)$};
\fill[red!85!black] (1,3) circle (3pt);
\node[red!75!black,anchor=west] at (1.15,3.25) {$p_\ell=(1,3)$};
\end{tikzpicture}%
}
\end{minipage}
\caption{The minimum cut $C$ is crossed by a link $\ell = (v_1, v_3)$. This link is represented by the point $(1,3) \in \mathbb{Z}^2$, which is contained in the union of the leftless rectangle $R^L(C)$ and the topless rectangle $R^T(C)$.}
\label{fig:cycle-geometric-reduction}
\end{figure}

Indeed, we name the vertices of $G$ in cyclic order as $V = (v_0,v_1,\ldots,v_{n-1},v_0)$ with an arbitrary vertex selected as the ``left-most" vertex $v_0$. Then a min-cut $C \subseteq V \setminus v_0$ can be represented as an \textit{interval} in the cyclic order $C=\{v_a,v_{a+1},\ldots,v_b\}, \text{where } 1\leq a\leq b\leq n-1.$
A link $\ell=(v_i,v_j)$  crosses the cut $C$ if it enters the interval through its left boundary or
leaves it through its right boundary. We can represent this with a pair of inequalities: for $\ell$ to enter $C$ from the left, we must have 
$i<a\leq j\leq b.$
Similarly, a link $\ell = (i,j)$ enters $C$ from the right exactly when $a\leq i\leq b<j$. Thus, the link $\ell = (v_i,v_j)$ crosses $C$ exactly when the point $(i,j)$ is contained in $S = R^L \cup R^T$, where 
\[
R^L(C)=(-\infty,a-1]\times[a,b] \text{ and } R^T(C)=[a,b]\times [b+1,\infty)
\]
are leftless and topless rectangles, respectively. Hence, for a cycle with $n$ vertices, we recover an instance of Online TL hitting set with at most $n^2$ points which lie on the $n \times n$ integer grid. See \Cref{fig:cycle-geometric-reduction} for an illustration of the reduction.

\subsubsection{Hitting TL-rectangles online}
It now suffices to provide competitive bounds for the Online TL-rectangle hitting set problem. 
The weighted case is our main technical contribution, and is much more involved than the unweighted case. Hence, we explain each separately, and start with the simpler unweighted setting.

\paragraph{Unweighted Online TL-rectangle Hitting} We mainly utilize a previously known logarithmic-competitive algorithm for hitting bottomless rectangles, given by De, Singh, and T\'oth~\cite{DeSinghToth2025}.  
By rotation, their result easily extends to hitting leftless rectangles, or to hitting topless rectangles. Can we leverage this algorithm to obtain a logarithmic-competitive algorithm which hits unions of such shapes? One natural idea is to try to run both algorithms in parallel, essentially hitting \textbf{both} the topless and leftless rectangle on each arrival, and hope that this blows up the competitive ratio by a factor of 2 at most. Unfortunately, this does not work in general, but we are able to use a stronger property of their analysis which allows us to handle unions. 

\begin{restatable}[\textcite{DeSinghToth2025}]{theorem}{toth}\label{thm:toth}
Let $P \subseteq [0,N)^2\cap \mathbb Z^2$ be a point set. Then there is a deterministic algorithm for online bottomless rectangle hitting set satisfying:
\begin{enumerate}
    \item\label{DST-primal} If a bottomless rectangle is satisfied on arrival, the algorithm adds no
    new points. Otherwise, it adds at most two points.

    \item\label{DST-dual} For every point $p \in P$, there are $O(\log N)$ rectangles containing $p$ that are unsatisfied on arrival.

\end{enumerate}
\end{restatable}

The above two properties easily imply an $O(\log N)$-competitive algorithm for hitting bottomless rectangles if the points lie on the $N \times N$ integer grid. Moreover, we are able to use the refined guarantee to design an $O(\log N)$-competitive algorithm for Online TL-rectangle hitting via a simple primal-dual fitting argument. 

\paragraph{Weighted Online TL-rectangle Hitting} Our algorithm for the weighted case is significantly more involved. We ultimately obtain an $O(\log n \log \log n)$-competitive algorithm for weighted online TL-rectangle hitting set with $n$ points. We provide a high level overview of the techniques here. 

Our main approach involves three-steps. First, as with any online hitting set problem with $n$ elements, we can use the classical algorithm of~\cite{AlonAABN09} to obtain an $O(\log n)$-competitive \textit{fractional} solution $x^t$. Our next goal is to design an online (randomized) rounding algorithm which takes as input an arbitrary increasing fractional solution $x^t$ at each time, and yields a feasible integral online solution. We show how to design such an online rounding algorithm for weighted bottomless rectangles with a loss of at most $O(\log \log n)$. Finally, our third step is a general observation that, given online rounding algorithms for two classes of shapes $\mathcal{S}_1$ and $\mathcal{S}_2$ with losses $\alpha_1(n)$ and $\alpha_2(n)$, we can construct an online rounding algorithm for the class consisting of unions of shapes from $\mathcal{S}_1$ and $\mathcal{S}_2$ with loss $2(\alpha_1(n) + \alpha_2(n))$. Applying this where $\mathcal{S}_1$ consists of leftless rectangles and $\mathcal{S}_2$ consists of topless rectangles yields an online rounding algorithm for TL-rectangles with loss $O(\log \log n)$. Thus, combining this with the $O(\log n)$-competitive fractional solution yields the desired bound.

Steps 1 and 3 are relatively straightforward, so we elaborate here on our online rounding algorithm for bottomless rectangles. In particular, we prove the following:

\begin{restatable}{lemma}{thm:rounding-loss}\label{lem:rounding}
Given an instance of Online Bottomless-Rectangle hitting set with $|P| = n$, arriving request sequence $R^t$, and an online fractionally feasible solution $x^t$, there is an online randomized rounding algorithm which outputs an increasing sequence of feasible integral hitting sets $H^t \subseteq P$ so that $\Pr[p \in H^t] \leq O(\log \log n) x^t_p$.  
\end{restatable}

To prove \Cref{lem:rounding}, we exploit a structural property that has been used to obtain improved algorithms for hitting geometric set systems.
The \emph{shallow cell complexity (SCC)} $\varphi(\ell,k)$ bounds the number of sets of size $k$ in any restriction of the set system to $\ell$ elements.  
It is known that Bottomless rectangles, in particular, have low SCC. The concept of SCC was first introduced by \textcite{ChanGKS12} to obtain improved approximations for geometric set systems in the \textit{offline} setting. First, they solve the natural LP relaxation to obtain a feasible fractional solution $x$. In classical randomized rounding, one would then sample elements with probability proportional to their value in the fractional solution. Here, they instead create a number of duplicates of each element $e$, where the number of copies of $e$ is proportional to $x_e$ (after zeroing elements of small $x$ value). They then employ a method called Quasi-uniform sampling which exploits the low-SCC of the instance to obtain random integral solution with improved bounds. 

There are two issues when trying to employ Quasi-uniform sampling in the online setting. First, it is unclear how to correlate the irrevocable samples in each time step to preserve the correct marginals. Second, since the values of $x^t$ are increasing, we need to use an increasing number of copies of each element over time. A recent paper of \textcite{bhore2026} attempts to handle these exact issues. In particular, they \textit{successfully} design an online sampler which incurs a low loss if the SCC of the duplicated set system remains low. But, their claim that the duplicated set system preserves low SCC is false, and the paper was subsequently retracted. 

However, we can get around this hurdle. The key is to define a fixed set system upfront with plenty of copies of each element. Then, as the value of $x_e$ increases over the course of the algorithm, we ``activate" an increasing number of copies of $e$, so that each arriving shape is hit only by activated copies. We then directly show that the SCC remains low for the resulting set family, and employ (a specialization of) the sampler of~\cite{bhore2026} to obtain our result.

\subsubsection{Unweighted OCycAP with node-arrivals}
It is striking that our methods can obtain optimal, or nearly optimal competitive ratios even in the more general setting of cut-arrivals. One might wonder whether there is any difference in complexity between the two arrival models. We show that, for the case of unweighted Online Cycle Augmentation, there is a separation.

First, let us explain why there is no $o(\log n)$-competitive algorithm for unweighted OCycAP for cut arrivals. In fact, this result already follows from our earlier result which establishes an equivalence between OCycAP and OCAP under the cut-arrival model. Since there is no $o(\log n)$-competitive algorithm for Online TAP~\cite{NaorUW22}, even in the unweighted setting, there cannot be such an algorithm for OCAP. Hence, by \Cref{thm:redtocycle}, there is no $o(\log n)$-competitive algorithm for OCycAP either. 

That being said, there is also a clean, direct lower bound which takes place directly on a cycle instance, which we prove in \Cref{sec:ocycap-cut-lb}. The lower bound also applies to fractional, and hence randomized algorithms as well. 

\paragraph{OCycAP Algorithm for Node Arrivals} In light of this, we turn to the node-arrival model. Here, we take advantage of the structure of the \emph{link intersection graph}. If $G$ is a cycle, then every link $\ell \in L$ can be viewed as a chord of the cycle. Informally, the link intersection graph is the graph whose node set corresponds to $L$, with an edge between $\ell$ and $\ell'$ if they intersect as chords. It is well-known~\cite{BasavarajuEtAl2014} in the node arrival setting that satisfying a terminal-pair $(u, v)$ is equivalent to purchasing a set of links containing a path in the link intersection graph from a link incident to $u$ to a link incident to $v$. 

The algorithm maintains the set of \emph{active} connected components of the link intersection graph that contain any terminals seen so far. When a terminal pair $(u, v)$ arrives, we iteratively expand the component containing $u$ by either choosing two links that merge it with another active component, or if no such choice exists, choosing two links that expand it farthest in both directions along the cycle. Using a careful charging argument, we can independently bound the total cost attributed to all merges, as well as the total cost of all steps in which a merge could not be performed, each by a constant times the cost of the optimal solution.

\subsection{Related Work}

\paragraph{Offline Connectivity Augmentation}
There is a great deal of work on Connectivity Augmentation in the offline setting. Focusing on the weighted case, the problem was first introduced by Eswaran and Tarjan~\cite{EswaranTarjan1976}, who showed WCAP is NP-complete. Later, a 2-approximation was given by Frederickson and J{\'a}J{\'a} for WTAP~\cite{FredericksonJaja1981}. After a long line of work on various special cases~\cite{CohenNutov2013, FioriniEtAl2018, TraubZenklusen2022WTAP}, Traub and Zenklusen gave a 1.5 + $\varepsilon$ approximation for offline WTAP, and then matched this for offline WCAP~\cite{TraubZenklusen2025WTAP, TraubZenklusen2023WCAP}. More recently, Cohen-Addad, Drygala, Klein, and Svensson, gave an $\alpha$-approximation for WTAP where $\alpha < 1.5$~\cite{CohenAddadDrygalaKleinSvensson2026}. After various improvements over the years~\cite{Adjiashvili2019, CheriyanGao2018a, CheriyanGao2018b, EvenFeldmanKortsarzNutov2009, GrandoniKalaitzisZenklusen2018}, the current best approximation factor for unweighted TAP and CAP is 1.393 due to Cecchetto, Traub and Zenklusen~\cite{CecchettoTraubZenklusen2021}.

\paragraph{Online Geometric Hitting Set} An instance of online Geometric Hitting Set is defined by a set $P$ of points in Euclidean space and a collection of shapes which arrive online to be hit. As mentioned, the algorithm of~\cite{AlonAABN09} can be applied to any such problem to obtain an $O(\log p \log m)$-competitive ratio if there are $p$ points and $m$ possible arriving sets. This holds even in the weighted case. However, improved competitive ratios can be achieved in certain  geometric settings. Even and Smorodinsky~\cite{EvenSmorodinsky2014} obtained optimal $O(\log p)$ competitive ratios
for the unweighted case when the arriving shapes are
are intervals on a line, or half-planes, or unit disks
in the plane. Khan, Lonkar, Rahul, Subramanian and Wiese~\cite{KhanEtAl2023}
gave an $O(\log N)$-competitive algorithm for
axis-aligned squares when the points lie on an
$N \times N$ integer grid, again in the unweighted setting. More recently, De, Singh, and
T\'oth~\cite{DeSinghToth2025} obtained an
$O(\log N)$-competitive algorithm for hitting bottomless
rectangles on an $N \times N$ integer grid, as well as an
$O(\log M \log p)$-competitive algorithm for disks
with radii in $[1,M]$. As far as we can tell, prior to this work, there are no improvements over the algorithm of \cite{AlonAABN09} for any weighted geometric special cases of Online Hitting Set, excluding the retracted paper of~\cite{bhore2026}.

\subsection{Organization of the paper}
In \Cref{sec:prelim}, we formalize the problem models, online hitting set formulations, and preliminaries. In \Cref{sec:OCAP}, we present the optimal $O(\log n)$-competitive algorithm for unweighted OCAP under cut-arrivals via reductions to cycles and TL-rectangles. In \Cref{sec:OWCAP}, we present the $O(\log n \log \log n)$-competitive randomized algorithm for weighted OCAP under cut-arrivals using our online rounding framework. Finally, in \Cref{sec:unweighted-ocycap}, we establish the separation on cycles by giving a $O(1)$-competitive algorithm for node-arrivals and an $\Omega(\log n)$ lower bound for cut-arrivals.

\section{Preliminaries}
\label{sec:prelim}

In all online connectivity augmentation models, we are given a $k$-edge-connected graph $G = (V, E)$ with $|V| = n$ and a set of candidate links $L \subseteq \binom{V}{2}$ upfront. The links have non-negative costs given by $c : L \to \mathbb{R}_{\ge 0}$. In the unweighted setting, we assume $c(\ell) = 1$ for all $\ell \in L$. 

For a subset of vertices $C \subseteq V$, we denote by $\delta_E(C)$ the set of edges in $E$ with exactly one endpoint in $C$, and by $\delta_L(C)$ the set of candidate links in $L$ with exactly one endpoint in $C$. Since $G$ is $k$-edge-connected, every cut $C$ has $|\delta_E(C)| \ge k$. A cut $C$ is a \emph{minimum cut} (or $k$-cut) if $|\delta_E(C)| = k$. A link $\ell \in L$ is said to \emph{cover} or \emph{cross} a cut $C$ if $\ell \in \delta_L(C)$. 

Requests arrive online over time in discrete time steps. We use $t \ge 1$ to denote the current time step. 

\subsection{Online Connectivity Augmentation Models}

Depending on the setting, requests either take the form of minimum cuts that need to be covered, or terminal vertex pairs that need to be connected.

\begin{problem}[OWCAP and OCAP with cut arrivals]
    A sequence of $k$-cuts $C_1, C_2, \ldots$ of $G$ arrive online one at a time. When a cut $C_t$ arrives at time step $t$, we must irrevocably purchase a subset of links to add to our solution. At the end of time step $t$, our cumulative purchased link set $F_t \subseteq L$ must satisfy $|\delta_{F_t}(C_i)| \ge 1$ for all $i \le t$. The objective in the Weighted Online Connectivity Augmentation Problem (OWCAP) is to minimize the total cost $\sum_{\ell \in F_t} c(\ell)$. In the unweighted version (OCAP), the goal is to minimize $|F_t|$.
\end{problem}

\begin{problem}[OWCycAP and OCycAP with cut arrivals]
    This is the special case of OWCAP (or OCAP) with cut arrivals in which the given base graph $G$ is a cycle and $k = 2$.
\end{problem}

\begin{problem}[OWCAP and OCAP with node arrivals]
    In each time step $t$, a pair of terminal vertices $(s_t, \tau_t) \in \binom{V}{2}$ arrives. We must irrevocably purchase links so that the vertices $s_t$ and $\tau_t$ are $(k+1)$-edge-connected in the augmented graph. That is, at the end of time step $t$, the cumulative purchased link set $F_t \subseteq L$ must be such that every cut $C$ separating $s_t$ and $\tau_t$ has $|\delta_{E \cup F_t}(C)| \ge k+1$. As before, OWCycAP and OCycAP denote the special cases where $G$ is a cycle.
\end{problem}

\subsection{Online Hitting Set and Competitive Analysis}

The Connectivity Augmentation problems defined above can be formulated as instances of the Online Hitting Set problem.

\begin{problem}[Online Hitting Set]
    Initially, we are given a set system $(U, \mathcal{F})$ where $U$ is a universe of $N$ elements with non-negative costs $c : U \to \mathbb{R}_{\ge 0}$ and $\mathcal{F}$ is a family of non-empty subsets of $U$. A sequence of requests $R_t \in \mathcal{F}$ arrive online. At each time step $t$, we must maintain an integral hitting set $H_t \subseteq U$ such that $H_t \cap R_{t'} \neq \emptyset$ for every $t' \le t$, and $H_t \supseteq H_{t-1}$. The goal is to minimize $\sum_{e \in H_t} c(e)$.
\end{problem}

\paragraph{Fractional Hitting Set and LP Formulation.}
In the fractional relaxation of the problem, we must maintain a monotone fractional solution $x^t \in \mathbb{R}^U_{\ge 0}$ over time. The problem can be expressed as an online covering linear program:
\begin{align*}
    \text{minimize} \quad & \sum_{e \in U} c(e) x^t_e \\
    \text{subject to} \quad & \sum_{e \in R_{t'}} x^t_e \ge 1 \quad \forall t' \le t, \\
    & x^t_e \ge x^{t-1}_e \quad \forall e \in U, \\
    & x^t_e \ge 0 \quad \forall e \in U.
\end{align*}

\paragraph{Competitive Analysis.}
The performance of our online algorithms is measured using competitive analysis. 
\begin{itemize}
    \item \textbf{Deterministic Competitive Ratio:} A deterministic online algorithm is $\alpha$-competitive if, for any valid request sequence, the cost of the algorithm's solution is at most $\alpha \cdot \text{OPT}$, where $\text{OPT}$ is the cost of the optimal offline solution that knows the entire request sequence in advance.
    \item \textbf{Randomized Competitive Ratio:} For randomized algorithms, we evaluate the performance against an \emph{oblivious adversary}, which has knowledge of the algorithm but cannot adapt to its random coin flips. An algorithm is $\alpha$-competitive in expectation if, for any request sequence generated by an oblivious adversary, $\mathbb{E}[\text{cost}(\text{ALG})] \le \alpha \cdot \text{OPT}$. 
    \item \textbf{Fractional Competitive Ratio:} An online fractional algorithm is $\alpha$-competitive if it maintains a valid fractional solution $x^t$ whose total cost is at most $\alpha$ times the cost of the optimal offline fractional solution.
\end{itemize}

\section[An O(log n)-competitive algorithm for OCAP under cut arrivals]{An $O(\log n)$-competitive algorithm for OCAP under cut arrivals}\label{sec:OCAP}

In this section, we give a simple $O(\log n)$ competitive algorithm for OCAP under cut arrivals.

\subsection{The cycle case}
\label{sec:ocycap}
We begin by giving an $O(\log n)$-competitive algorithm for OCycAP on a cycle with cut arrivals. Later, we show how to use this to handle the case of a general graph.

\begin{problem}[OCycAP with cut arrivals]
    Upfront, there is a cycle $G = (V,E)$ and a collection of links $L \subseteq \binom{V}{2}$. At time step $t$, a minimum cut $C_t$ arrives to be augmented. We seek to maintain a set of links $F_t \subseteq L$ of minimum cardinality so that $F_t$ crosses all min-cuts which have arrived by time $t$. 
\end{problem}

We now reduce the problem to an online geometric hitting set problem where the sets to be hit correspond to cuts, and points correspond to links. In particular, we convert each minimum cut $C$ into the union of a leftless rectangle and a topless rectangle as follows. 
Order the vertices of $G$ as $V=(v_0,v_1,\ldots,v_{n-1},v_0)$ along the cycle, with an arbitrary vertex selected as $v_0$. A minimum cut $C$ separates the cycle into two paths; we choose the path
that does not contain $v_0$ to represent the cut:
$C=\{v_a,v_{a+1},\ldots,v_b\}, \text{where } 1\leq a\leq b\leq n-1.$
We refer to this cut as $C = [a,b]$. For a link $\ell=(v_i,v_j)$ with $i<j$, the link crosses the cut $C = [a,b]$ precisely when it enters the interval through its left boundary or
leaves it through its right boundary. For the link $\ell$ to enter from the left, we must have 
$i<a\leq j\leq b.$
Similarly, the link $\ell = (i,j)$ enters $C = [a,b]$ from the right exactly when $a\leq i\leq b<j.$

We can now encode this as a geometric hitting set problem as follows. For each link $\ell = (v_i,v_j)$, we create a point $p_\ell = (i,j) \in \mathbb{Z}^2$. For each arriving cut $C = [a,b]$, we create two rectangles which are subsets of $\mathbb{R}^2$:

\[
R^L(a,b)=(-\infty,a-1]\times[a,b]
\]
and
\[
R^T(a,b)=[a,b]\times[b+1,\infty).
\]

Notice that $R^L$ is a leftless rectangle and $R^T$ is a topless rectangle, and all corners lie on the integer lattice.

\begin{proposition}\label{prop:geometric-covering}
    A link $\ell = (v_i,v_j)$ with $i < j$ crosses the minimum cut $C = [a,b]$ if and only if $p_\ell = (i,j) \in R^L(a,b) \cup R^T(a,b)$.
\end{proposition}

Hence, we have reduced the problem to Online TL-rectangle hitting set, \Cref{prob:TL-hitting-set}. We now show how to leverage a recent result on hitting set for bottomless rectangles to yield an $O(\log n)$ competitive algorithm for the problem. We use a stronger property of the analysis which allows us to handle unions.

\toth*

The above two properties imply an $O(\log N)$-competitive algorithm for bottomless rectangle hitting set where $P \subseteq [0,N)^2\cap \mathbb Z^2$. By rotation, this algorithm can easily be adapted to handle topless and leftless rectangles with the same guarantee. We now give the $O(\log n)$-competitive algorithm for OCycAP. See~\Cref{alg:cut-arrivals}.

\begin{algorithm}[H]
\caption{OCycAP under cut arrivals}\label{alg:cut-arrivals}

\textbf{Input:} A cycle $G=(V,E)$ where $|V|=n$, a set of links
$L\subseteq \binom{V}{2}$, and an online sequence of minimum cuts
$C_1,\ldots,C_m$.

Let
\[
P=\{(i,j): (v_i,v_j)\in L,\ i<j\}
    \subseteq [0,n)^2\cap\mathbb Z^2.
\]

Initialize $F\gets\emptyset$.

Initialize a leftless-rectangle hitting-set algorithm $\mathcal A^L$ on $P$

Initialize a topless-rectangle hitting-set algorithm $\mathcal A^T$ on $P$.

\textbf{For} each arriving cut $C=[a,b]$:

\quad If $C$ is already covered by $F$, do nothing.

\quad If $C$ is not covered by $F$:

\qquad Let
\[
R^L=(-\infty,a - 1]\times[a,b]
\qquad\text{and}\qquad
R^T=[a,b]\times[b + 1,\infty).
\]

\qquad Feed the leftless rectangle $R^L$ to the algorithm $\mathcal A^L$ and let $Q^L$ be the set of points purchased.

\qquad Feed the topless rectangle $R^T$ to the algorithm $\mathcal A^T$ and let $Q^T$ be the set of points purchased.

\qquad Add to $F$ the links corresponding to points in $Q^L\cup Q^T$.

\end{algorithm}

\Cref{prop:geometric-covering} essentially shows that OCycAP with cut arrivals is equivalent to an online geometric hitting set problem where each arriving shape is the union of a topless and a leftless rectangle. \Cref{alg:cut-arrivals} handles the union by purchasing points to cover \textit{both the topless and the leftless rectangle, as long as the arriving union shape is not already hit}. 

We now show that this yields the desired competitive guarantee. 

\begin{theorem}
    \Cref{alg:cut-arrivals} is an $O(\log n)$-competitive algorithm for OCycAP with cut arrivals. 
\end{theorem}
\begin{proof}
We will use a dual-fitting argument. First we describe the covering LP for OCAP and its dual. 
\begin{center}
\begin{minipage}{0.47\textwidth}
\centering
\[
\begin{array}{ll}
\text{minimize}
    & \displaystyle \sum_{\ell\in L} x_\ell \\[1ex]
\text{subject to}
    & \displaystyle \sum_{\ell:\,\ell\text{ covers }C_t} x_\ell \geq 1
      \quad \forall t, \\[2ex]
    & x_\ell \geq 0
      \quad \forall \ell\in L.
\end{array}
\]
\end{minipage}
\hfill
\begin{minipage}{0.47\textwidth}
\centering

\[
\begin{array}{ll}
\text{maximize}
    & \displaystyle \sum_{t=1}^k y_t \\[1ex]
\text{subject to}
    & \displaystyle \sum_{t:\,\ell\text{ covers }C_t} y_t \leq 1
      \quad \forall \ell\in L, \\[2ex]
    & y_t \geq 0
      \quad \forall t.
\end{array}
\]
\end{minipage}
\end{center}

Consider a run of Algorithm~\ref{alg:cut-arrivals}. We first construct
a primal solution from the run of the algorithm. For each link
$\ell\in L$, set $x_\ell=1$ if $\ell$ is purchased by the algorithm,
and set $x_\ell=0$ otherwise. So $x$ is a feasible primal
solution with  objective value $|F|$.

We next construct a dual solution. Let $T$ denote the set of iterations
at which the arriving cut is not already covered by $F$. Define
$y_t=1$ if $t\in T$, and $y_t=0$ otherwise. Thus the value of the dual
solution is $|T|$. Also, $y$ is $O(\log n)$-approximately feasible. Indeed, by property \ref{DST-dual} of \Cref{thm:toth}, for every link \(\ell\), the corresponding point \(p_\ell\) is contained in at most \(O(\log n)\) unsatisfied rectangles in each of the two geometric instances. Therefore \(\ell\) covers at most \(O(\log n)\) cuts \(C_t\) for which \(y_t=1\). So there exists a constant $c$ such that 
$\sum_{t:\,\ell\text{ covers }C_t} y_t \leq c\log n$ for every link $\ell \in L$. By weak duality, we conclude that $\OPT \geq |T| / c\log n$.

Finally, by definition of \Cref{alg:cut-arrivals}, it purchases no points on iterations outside of $T$, and by property \ref{DST-primal} of \Cref{thm:toth}, it purchases at most four links for each iteration in \(T\). Hence \(|F|\leq 4|T| \leq (4c\log n) \cdot \OPT \). Therefore, \Cref{alg:cut-arrivals} is $O(\log n)$-competitive.
\end{proof}

\subsection{From OCycAP to general OCAP}

We now show how to reduce the OCAP with cut arrivals problem on a general graph $G$, to the same problem where $G$ is a cycle. This is a simple corollary of the following theorem, due to Dinitz, Karzanov, and Lomonosov~\cite{DinitzKarzanovLomonosov1976} (called ``Theorem on circumference disposition'' in their paper).

\begin{theorem}[Dinitz, Karzanov, and Lomonosov \cite{DinitzKarzanovLomonosov1976}]
\label{thm:DKL}
    Given any connected graph $G = (V,E)$, there exists an (efficiently computable) cycle $H = (V,E')$ such that for all $C \subseteq V$, if $\delta_E(C)$ is a minimum cut in $G$, then $\delta_{E'}(C)$ is a minimum cut in $H$.
\end{theorem}

\begin{proposition}
\label{prop:OCAP-to-OCycAP}
    If there is an $\alpha$-competitive algorithm $\mathcal{A}$ for OCAP with cut-arrivals on a cycle, then there is an $\alpha$-competitive algorithm for OCAP with cut-arrivals on a general graph. This holds in the weighted case as well.
\end{proposition}
\begin{proof}
    Given an instance of OCAP with online cut-arrivals $C_1, \ldots, C_k$ in a general graph $G = (V,E)$, we apply the reduction given in \Cref{thm:DKL} to obtain a cycle $H = (V,E')$. The link set $L$ considered as a subset of $\binom{V}{2}$ is exactly the same. If the links have costs, these are preserved as well.

    Now, whenever a minimum cut $C_t$ arrives to be augmented in $G$, we will cause the minimum cut $C_t$ to arrive in $H$. The set of links which hit $C_t$ in $G$ is identical to the set of links which hit $C_t$ in $H$. The algorithm $\mathcal{A}$ will maintain a set of links $F$ which hits all cuts which have arrived so far in $H$. This set of links $F$ is also feasible for $G$. Furthermore, the optimal solutions for covering $C_1,\ldots,C_k$ in $G$ and $H$ are equal. Thus the set of links $F$ is $\alpha$-competitive for $G$ as well.
\end{proof}

\section[An O(log(n)loglog(n))-competitive algorithm for OWCAP under cut arrivals]{An $O(\log n\log\log n)$-competitive algorithm for OWCAP under cut arrivals}\label{sec:OWCAP}

Recall that OWCAP reduces to OWCycAP (\Cref{thm:redtocycle}). Also OWCycAP  on cycle with $n$ vertices reduces to online TL-rectangle hitting set (\Cref{thm:cycletoTL}) with $O(n^2)$ points. Thus, to prove \Cref{thm:OWCAP}, it suffices to give a randomized online algorithm for online TL-rectangle hitting set with expected competitive ratio $O(\log n \log \log n)$, where $n$ now refers to the number of points in the instance. Our approach is to obtain an online fractional hitting set and then round it. Since there is an $O(\log n)$-competitive algorithm for online fractional hitting set in arbitrary set systems \cite{AlonAABN09}, it remains to design an online rounding algorithm for fractional TL-rectangle hitting set with a $O(\log \log n)$ overhead. Thus, the goal in this section is to prove the following theorem.

\begin{theorem}
    \label{thm:TL-rounding}
    If there exists a $\rho(n)$-competitive fractional algorithm for online TL-rectangle hitting set, then there exists a randomized online algorithm for online TL-rectangle hitting set with expected competitive ratio $O(\rho(n)\log \log n)$.
\end{theorem}

We prove the theorem in two steps. Recall that in online TL-rectangle hitting set, the requests are unions of topless and leftless rectangles. We first show how to round an online fractional solution for rectangles that are unbounded in a single direction (we canonically use bottomless rectangles for this section). In \Cref{sec:online-rounding-bottomless}, we give a rounding algorithm such that at every time step $t$, the point $p$ is included in its hitting set with probability at most $O(\log \log n)x^t_p$. This is the crux of our approach. Then in \Cref{sec:union-rounding}, we show how to use rounding algorithms for topless and leftless rectangles to obtain a rounding algorithm for their union.
We give intuitive explanations of these steps below.

Once we have a fractional solution, classical randomized rounding requires us to boost probabilities by a factor of $O(\log n)$ to ensure that the sampled solution is a feasible hitting set. But this loss is more than what we can accommodate. Instead we use an online sampler that exploits properties of the range space (namely its shallow cell complexity), to provide a tighter bound. 

The online sampler is based on the Quasi-uniform sampling approach developed by~\cite{ChanGKS12} for the offline setting. To extend this to the online setting, we would ideally create a new point set, which consists of copies of every point in the input, where the number of copies is proportional to the value assigned to that point by the fractional solution. But since the underlying universe of points cannot keep changing in the online setting, we create $n$ copies for each point upfront, but only \textit{activate} an appropriate subset that depends on the fractional value assigned to the point at any given time step.
To show that the range system has bounded shallow cell complexity, we imagine the set of points to be in 3D space, with the third dimension being their time of activation. Finally, we show that the arriving sets can be thought of as octants, which can be shown to have the required shallow cell complexity.

To handle shapes which are the union of a topless and a leftless rectangle, we do not deal with the geometry but work directly with the fractional solution, and modify the rounding step. In particular, since $x$ is fractionally feasible, one of the two constituent rectangles is hit to an extent at least $1/2$. So, we consider only the heavier rectangle, double the values assigned to these points (to maintain feasibility), and work with this new fractional solution to do the rounding step. This loses a factor of 2 at most, so we obtain the same $O(\log \log n)$ bound for the union. 

\subsection{Online rounding for bottomless rectangles}
\label{sec:online-rounding-bottomless}

We first introduce some necessary terminology on shallow cell complexity. 

\begin{definition}[Traces and shallow cell complexity]
\label{def:scc}
Let $(\mathcal{V}, \mathcal{G})$ be a set system on a finite universe of elements $\mathcal{V}$. 
\begin{enumerate}
    \item \textbf{Trace:} For any subset $W \subseteq \mathcal{V}$, the \emph{trace} of $\mathcal{G}$ on $W$ is the family of induced subsets $\mathcal{G}|_W := \{ Q \cap W : Q \in \mathcal{G} \}.$
    \item \textbf{Shallow trace:} For an integer $\kappa \ge 1$, a trace $S \in \mathcal{G}|_W$ is called a \emph{$\kappa$-shallow} trace if we have $1 \le |S| \le \kappa$.
    \item \textbf{Shallow cell complexity (SCC):}  We say that $(\mathcal{V}, \mathcal{G})$ has shallow cell complexity $\varphi$ (for function $\varphi : \mathbb{N} \times \mathbb{N} \to \mathbb{R}_{\ge 0}$), if for every subset $W \subseteq \mathcal{V}$ with $|W| = m$,
    \[
        \big| \{ S \in \mathcal{G}|_W : 1 \le |S| \le \kappa \} \big| \le m \cdot \varphi(m, \kappa).
    \]
    In other words,  $(\mathcal{V}, \mathcal{G})$ has an SCC of $\varphi$ if the number of $\kappa$-shallow traces on any set of size $m$ is at most $m \cdot \varphi(m,\kappa)$.
    
\end{enumerate}
\end{definition}

Fix \(P \subseteq \mathbb{R}^2\), a fixed universe of \(n\ge1\) points
with non-negative 
costs \(c_p\). Requests $R_t$ arrive online and the goal is to maintain an online hitting set $H$ whose total $c(H) = \sum_{p \in H} c_p$ is minimized. We first focus on the case where each request $R_t$ is a bottomless rectangle. Let $x$ be an online fractional solution. 

Denote by $x^t_p$ the extent to which point $p$ is bought in response to the first $t$ requests, and define $x^t(S) = \sum_{p \in P \cap S} x^t_p$.

\begin{theorem}
    \label{thm:bottomless-rounding}
    There exists a randomized algorithm that given an online sequence of bottomless rectangles and an online fractional solution $x$, maintains an online hitting set $H$ such that $\Pr[p \in H] \leq O(\log \log n) x_p$ for every $p \in P$. The request sequence and $x$ are independent of the algorithm's randomness.
\end{theorem}

\subsubsection{Online sampling for bounded SCC set systems}
Once we have a fractional solution, we will use a specialized form of randomized rounding to obtain an integral solution. Rather than sampling with probabilities simply boosted from the fractional solution as in the classical randomized rounding approach, we use the following online sampler which exploits the geometric properties of our set system. The property used here is that the sets that arrive have a low SCC. The existence of such a sampler follows from a result of Bhore, Gupta and Kumar~\cite{bhore2026}. We note that, as of October 8, 2026,~\cite{bhore2026} was retracted from arXiv due to an error in Theorem 3.2.  However, Theorem 5.5 in~\cite{bhore2026} (version 1) is logically separate from the retracted result, and still holds. We state a specialization of this theorem, which is sufficient for our setting.

\begin{lemma}[Low SCC Online Sampler]\label{lem:sampler}
Let $\mathcal{V}$ be a universe of size $N$, and let
$N^{1/3} \le B \le N$ and $1 \le A \le C_A \log N$
be parameters (for a universal constant $C_A$, to be determined later). Let $Q_1,Q_2,\ldots$ be an oblivious online hitting set request
sequence satisfying $|Q_t| \ge B$ for every $t$. Let $\mathcal{G} = \{Q_1,Q_2,\ldots\}$, and assume that the
set system $(\mathcal{V},\mathcal{G})$ has shallow cell complexity bounded by $$\varphi(m,\kappa) \le A\kappa^2$$ for all $m,\kappa \ge 1$. Then there exists a randomized online algorithm maintaining an
increasing hitting set $J_t \subseteq \mathcal{V}$ such that
\[
    \Pr[e \in J_t]
    \le O\left(\frac{\log(A\log N)}{B}\right)
\]
for every element $e \in \mathcal{V}$ and every finite horizon $t$.
\end{lemma}

\begin{proof}
    This lemma is a special case of \cite[Theorem~5.5]{bhore2026}, rephrased for our purposes. In particular, if we write the set family of requests as $\mathcal{G} = \{Q_1, Q_2, \dots\}$, then \cite[Theorem~5.5]{bhore2026} provides an online algorithm that maintains an increasing hitting set $J_t \subseteq \mathcal{V}$ with 
    \begin{equation}\label{eq:bhore-bound}
        \Pr[e \in J_t] \leq O\left(\frac{1}{B} \log\Big(B_{L^*} \cdot \log N \cdot \varphi_{\mathcal{G}}(N_{L^*}, B_{L^*}) \Big)  \right)
    \end{equation}
    where $B$ is a parameter satisfying $B > c_1$ for constant $c_1$ and for which $\abs{Q_t} \geq B$ for every $Q_t \in \mathcal{G}$ (these two conditions are clearly satisfied by the conditions of the lemma). $B_\ell$, $N_\ell$ and $L^*$ are defined as follows. 
    \[
        N_\ell := \frac{c_1 N}{2^\ell} \qquad B_\ell := \frac{B}{2^\ell}
    \]
    and $L^*$ is the largest $\ell \in [0, \log(\nicefrac{B}{c_1})]$ for which 
    \[
        h_{\mathcal{G}}(N_\ell, B_\ell) := \sqrt{\frac{8 \ln(B_\ell \ceil{\log N_\ell} \varphi_{\mathcal{G}}(N_\ell, B_\ell))}{B_\ell}} \leq \frac{1}{2}.
    \]
    Finally, $\varphi_\mathcal{G}$ must be a \emph{well-behaved} bound on the \emph{shallow cell complexity} of the set system $\mathcal{G}$. The condition in the statement of our lemma is that $\varphi_{\mathcal{G}}(\cdot, \kappa) \leq A\kappa^2$. And it is easy to verify that this bound is $(\nicefrac{3}{2}, 8)$-well-behaved in the sense of \cite[Definition 2.2]{bhore2026}, since for any $\kappa \geq 8$ we have $A\kappa^2 \leq (A(\nicefrac{\kappa}{2})^2)^{3/2}$. Thus, all hypotheses of \cite[Theorem~5.5]{bhore2026} are satisfied. 

    Since $B \geq N^{1/3}$, we can choose some $\hat{\ell} \in [0, \log(\nicefrac{B}{c_1})]$ so that $B_{\hat{\ell}} = \Theta(\log N)$. Observe that $N_{\hat{\ell}} \leq O(N)$, and since $A \leq O(\log N)$, we get
    \[
        h_{\mathcal{G}}(N_{\hat \ell}, B_{\hat \ell}) \leq \sqrt{\frac{8 \ln(B_{\hat\ell}^3 \cdot O(\log N) \cdot A)}{B_{\hat \ell}}} \leq \sqrt{\frac{O(\log \log N)}{\Theta(\log N)}} = o(1).
    \]
    In particular, for this choice of $\hat \ell$, we have $h_{\mathcal{G}}(N_{\hat \ell}, B_{\hat \ell}) \leq \nicefrac{1}{2}$ (for sufficiently large $N$), and hence $L^* \geq \hat{\ell}$ and $B_{L^*} \leq O(\log N)$. 

    Finally, making the substitutions for $\varphi_{\mathcal{G}}$ and $B_{L^*}$ in \eqref{eq:bhore-bound} yields the desired result.
\end{proof}

\subsubsection{Virtual online hitting set instance for sampler}
Armed with the online sampler from \Cref{lem:sampler}, we wish to use it to sample points with probability proportional to the values assigned to them by the fractional solution. We now describe how to duplicate the point set to simulate this sampling.

\paragraph{New point set:}
We refer to the given point set $P$ as the physical points, and construct a new universe of virtual points $\mathcal{P} = P \times [n]$, consisting of $n$ virtual copies for each physical point $p \in P$ (yielding $N = |\mathcal{P}| = n^2$). We denote by $\pi: \mathcal{P} \rightarrow P$ the projection mapping each virtual point $e = (p, i)$ to its physical parent $p$, and the corresponding inverse map $\pi^{-1}:P\rightarrow\mathcal{P}$ outputs the set of virtual copies for any physical point $p\in P$.
But, not all copies of a point $p\in P$ are to be used at a given time step $t$. Since we require the number of copies of $p$ to be proportional to its fractional cost $x^t_p$, we only \textit{activate} a fraction of the copies at time $t$. This is done as follows.

First we define the vector
\begin{equation}\label{eq:truncation}
 z^t_p=
 \begin{cases}
  0,&x^t_p<1/(2n),\\
  \min\{1,2x^t_p\},&x^t_p\ge1/(2n).
 \end{cases}
\end{equation}
and similar to the vector $x^t$, we also define $z^t(S) = \sum_{p \in P \cap S} z^t_p$.
Now, we say that at time step $t$ the first \(a^t_p=\lceil{nz^t_p}\rceil\) copies of point \(p\in P\) are active, i.e., $A_t := \bigcup_{p \in P} (\{p\} \times [a^t_p]) \subseteq \mathcal{P}$ is the set of all active copies at time $t$. For any subset $S \subseteq \R^2$, let $A_t(S) := A_t \cap \pi^{-1}(S)$ denote the set of active copies of points in $S$ at time step $t$.

We now show that the fractional solution $z^t$ is monotone, feasible, and not too large compared to the original fractional solution $x^t$.
\begin{lemma}\label{lem:copies}
The vector \(z^t\) is monotone and feasible, the sets $A_t$ are
increasing ($A_1\subseteq A_2 \subseteq\cdots $), and for any request $R_t$ and every point $p \in P$,
\[
 |A_t(R_t)|\ge n,\qquad
 |A_t\cap\pi^{-1}(p)|=a^t_p\le4n\,x_p^t.
\]
\end{lemma}
\begin{proof}
    If there were a single point $p'$ with $x^t_{p'}\ge 1/2$ in a satisfied query set $R_t$, then $z^t_{p'}\ge 1$, hence also satisfying $R_t$. Now, we note that we round down a variable $x^t_p$ only if its value is less than $1/(2n)$. Since there are $n$ variables, a value of at most $1/2$ gets discarded, giving us that $R_t$ still has value at least $1/2$, with each variable having value at most $1/2$. Since the new vector doubles the value of all such variables, we would still have $z^t(R_t)\ge 1$, making it a feasible solution.
    Similarly, since $x$ is monotone, and $z$ doubles the variables which satisfy the threshold of $\ge 1/(2n)$, $z$ is also monotone.

    When $z_p^t=0$, the corresponding $x_p^t\ge0$, so we trivially have $a^t_p\le4n\,x^t_p$. And, for every positive $z^t_p$, we have $nz^t_p\ge1$. Therefore
    \begin{equation}
    \label{eq:active-copies}
     a^t_p = \ceil{nz_p^t}\le2nz_p^t\le4n\,x_p^t.
    \end{equation}
    
    Finally,
    \[
     |A_t(R_t)|=\sum_{p\in P\cap R_t}\ceil{nz_p^t}
     \ge n\,z^t(R_t)\ge n. \qedhere
    \]
\end{proof}

\paragraph{Virtual Requests:}
Similarly we need to ensure that the sets being sent to the sampler belong to a family $\mathcal{F}$ satisfying the properties required for \Cref{lem:sampler}.
The sets are created in an online fashion, and when a physical request $R_t$ arrives, we translate this to a virtual request $Q_t\subseteq\mathcal{P}$ to be passed to the online sampler.

Ideally, we would like to send $A_t(R_t)$ as the virtual request. However, in general, this family of sets does not satisfy the low shallow cell complexity condition required to apply \Cref{lem:sampler}. Instead, we choose a subset of $A_t(R_t)$ that is sufficiently large and falls within a more structured family of sets (note that hitting a subset of the required set is sufficient to hit the original set).

Before we go into details of the translation, we give a brief introduction to \emph{range trees}, which are a structure used in our construction. Given a set of points on the line, we will construct a data structure that can represent any possible line segment on this set of points. We first start with the sorted list of points (including $-\infty$ and $\infty$), and consider the segments between consecutive points (called canonical intervals) to be leaves of a tree. On this set as leaves, we construct a balanced binary tree. Each node in this tree represents an interval, which is the union of all the canonical intervals stored in its subtree. And the height of this tree is $O(\log n)$, where $n$ is the number of points in the input set.

With the notion of range trees in hand, the translation procedure works as follows:
Consider the entire point set $P$ projected to the x-axis and construct a range tree on this set of points. For a node $u$ of the range tree, denote the set of points in $P$ whose projection lies in the interval corresponding to $u$ as $S_u\subseteq P$. 

When a bottomless rectangle request $R_t$ of the form $[a_t,b_t]\times(-\infty,c_t]$ arrives, we consider the deepest node $v$ in the range tree that contains the interval $[a_t,b_t]$, i.e., the node in the range tree such that its corresponding interval contains the range $[a_t,b_t]$, but none of its children's intervals do (or $v$ is a leaf).
W.l.o.g., we can assume that $a_t, b_t$ are $x$-projections of points in $P$.
Among the set of points in $S_v$, select the points that satisfy $y\le c_t$ (i.e., the set $R_t\cap P$), and consider the set of their active copies $A_t(R_t)$. If $v$ is a leaf, we simply use $A_t(R_t)$ directly. Otherwise, letting the two child nodes of $v$ be $v_\ell$ and $v_r$, we consider the sets $R_t\cap P\cap S_{v_\ell}$ and $R_t\cap P\cap S_{v_r}$, construct the corresponding sets of active copies $A_t^{\mathrm{left}}, A_t^{\mathrm{right}}$, and pick the one with larger number of points.

With the instance as described above, we can now give the rounding algorithm using this instance, and the online sampler:
\begin{algorithm}[ht]
\caption{Online Rounding for Bottomless Rectangles}
\label{alg:bottomless}
\textbf{Input:} A fixed point set $P \subseteq \mathbb{R}^2$, an arriving sequence of bottomless rectangles $R_t$, and an $O(\log n)$-competitive online fractional solution $x^t$.

\textbf{Output:} A monotonically increasing hitting set $H$, satisfying all requests up until the current time step.

\SetKwInOut{Init}{Initialize}
\Init{Hitting set $H \leftarrow \emptyset$\;
Construct 1D binary range tree on the horizontal coordinates of $P$\;
Initialize the online sampler (\Cref{lem:sampler}) on universe $\mathcal{P} = P \times [n]$ with parameters $N = n^2$, $B = \lceil n/2 \rceil$, and $A = \max\{1, 2C_{\mathrm{dom}}(\lceil \log_2 n \rceil + 1)\}$\;
}
\ForEach{arrival of bottomless rectangle $R_t = [a_t, b_t] \times (-\infty, c_t]$ with fractional vector $x^t$}
{
    \tcp{Step 1: Activate copies}
    \ForEach{$p \in P$}{
        Let $z^t_p \leftarrow \begin{cases} 0, & x^t_p < 1/(2n) \\ \min\{1, 2x^t_p\}, & x^t_p \ge 1/(2n) \end{cases}$\;
        Mark the first $a^t_p \leftarrow \lceil n z^t_p \rceil$ copies of $p$ in $\mathcal{P}$ as active\; 
    }
    Update $A_t \subseteq \mathcal{P}$ to be the set of all currently active copies\;

    \tcp{Step 2: Query decomposition via horizontal range tree}
    Find the deepest node $v$ of the range tree containing $[a_t, b_t]$\;
    \eIf{$v$ is a leaf}{
        $Q_t \leftarrow A_t(R_t)$\;
    }
    {
        Partition $A_t(R_t)$ into subsets $A_t^{\mathrm{left}}$ and $A_t^{\mathrm{right}}$ associated with $v$'s children\;
        Retain the heavier piece: $Q_t \leftarrow \arg\max \{|A_t^{\mathrm{left}}|, |A_t^{\mathrm{right}}|\}$\;
    }

    \tcp{Step 3: Online sampling}
    Feed $Q_t$ to the online sampler (\Cref{lem:sampler}) and update the sampled hitting set $J_t \subseteq \mathcal{P}$\;

    \tcp{Step 4: Purchase points}
    Purchase physical parents of sampled copies: $H \leftarrow H \cup \{\pi(e) : e \in J_t\}$\;
}
\end{algorithm}

\subsubsection{Bounding the SCC of the Virtual Instance}

\begin{definition}[Activation time]
\label{def:spacetime}
For each copy $e = (p, i) \in \mathcal{P}$, its \emph{activation time} $\tau_e$ is the first time step at which it becomes active, i.e., $\tau_e = \min\{t \ge 1 : e \in A_t\}$, with $\tau_e = T + 1$ if $e$ remains inactive at the final time step $T$.
\end{definition}
For the sake of our analysis, we map each copy $e\in\mathcal{P}$ of a physical point $p\in P$ with spatial coordinates $(u_p, v_p)$ to the point $(u_p, v_p, \tau_e)\in\mathbb{R}^3$.

We can show that the sequence of queries we send to the sampler have a small number of shallow traces.
\begin{lemma}[Shallow traces] \label{lem:shallow-traces}
    Let $\mathcal{F}$ denote the family of all candidate query sets $Q_t \subseteq \mathcal{P}$ that can be generated across all nodes of the range tree. For any subset $W\subseteq \mathcal{P}$ and integer $\kappa \ge 1$,
    \begin{equation}\label{eq:few-small-traces}
     \left|\{Q_t\cap W: Q_t\in\mathcal{F} \text{ and } 1\le |Q_t\cap W|\le \kappa\}\right|
     \le O(\log n\cdot|W|\cdot \kappa^2).
    \end{equation}
\end{lemma}

To prove the above, we first show an accessory lemma that is implied from earlier work by Afshani and Tsakalidis \cite{AfshaniT14}. A short proof is given in Appendix~\ref{app:octant-traces} for completeness.
\begin{theorem}[3D octant shallow traces \cite{AfshaniT14}]
    \label{thm:octant-traces}
    There exists an absolute constant $C_{\mathrm{dom}}$ such that for any $r$-element point set in $\mathbb{R}^3$, the number of distinct nonempty traces of size at most $\kappa$ under octants of a fixed orientation is at most $C_{\mathrm{dom}} r \kappa^2$, for every integer $\kappa \ge 1$.
\end{theorem}

Now we may proceed with the proof of \Cref{lem:shallow-traces}.

\begin{proof}[Proof of \Cref{lem:shallow-traces}]
    For the sake of analysis, let us project the set of virtual points $\mathcal{P}$ to $\R^3$ as follows: for a point $p=(u_p,v_p)\in P$, and for each copy $e = (p, i) \in \mathcal{P}$, we consider the projection of the copy to be the point $(u_p, v_p, \tau_e)$, where $\tau_e$ is the activation time for copy $e$ (so that $s_z \le t \iff e \in A_t$).
    We will show that the request sets that are sent to the sampler can be thought of as octants in $\R^3$ (locally on the range tree), and since we can bound the number of shallow traces for octants, we will be able to bound the number of shallow traces for the overall family of queries as well.

    Consider the set of points $\pi^{-1}(S_u)\subseteq\mathcal{P}$ contained within the x-interval $[a,b]$ represented by a node $u$ of the range tree. Consider any rectangle query $R_t = [a_t,b_t] \times (-\infty,c_t]$ arriving at time $t$ that we assign to node $u$ for query decomposition. We distinguish two cases:
    \begin{itemize}
        \item \textbf{If $u$ is an internal node:} Let the left and right child nodes of $u$, $u_\ell, u_r$ respectively, correspond to the intervals $[a,c]$ and $[c,b]$ respectively, where $[a_t,b_t]$ is completely contained in $[a,b]$, but not in $[a,c]$ or $[c,b]$ (i.e., $a_t \le c \le b_t$). Since all points in $S_{u_\ell}$ already satisfy $x \le c$ and all points in $S_{u_r}$ satisfy $x \ge c$, the candidate sets $A_t^{\mathrm{left}}$ and $A_t^{\mathrm{right}}$ corresponding to $R_t$ are given by
        \[
            T_\text{left}=\{s=(s_x, s_y,s_z)\in \pi^{-1}(S_{u_\ell}) \mid s_x \ge a_t \text{ and } s_y \le c_t \text{ and } s_z \le t\}
        \]
        and
        \[
            T_\text{right}=\{s=(s_x, s_y,s_z)\in \pi^{-1}(S_{u_r}) \mid s_x \le b_t \text{ and } s_y \le c_t \text{ and } s_z \le t\}.
        \]
        Hence, every query set generated from an internal node $u$ is an octant on the points $\pi^{-1}(S_u)$ projected to $\R^3$. The family of queries generated from this node is $\mathcal{F}_u=\mathcal{F}_\text{left}\cup\mathcal{F}_\text{right}$, which is the union of octants in two different orientations.
        \item \textbf{If $u$ is a leaf:} The algorithm directly outputs $Q_t = A_t(R_t)$. Since $[a_t, b_t] \subseteq [a,b]$, this query is represented by $T_\text{leaf} = \{s=(s_x, s_y, s_z) \in \pi^{-1}(S_u) \mid s_x \ge a_t \text{ and } s_y \le c_t \text{ and } s_z \le t\}$, which is an octant of a single orientation. Thus, $\mathcal{F}_u$ consists of octants in one direction.
    \end{itemize}

    Now, using \Cref{thm:octant-traces}, the number of shallow traces for the family of queries $\mathcal{F}_u$ generated from node $u$ is bounded by
    \begin{align*}
     |\{Q_t &\cap W: Q_t\in\mathcal{F}_u \text{ and } 1\le |Q_t\cap W|\le \kappa\}| \\
     &\le |\{Q_t\cap W: Q_t\in\mathcal{F}_\text{left} \text{ and } 1\le |Q_t\cap W|\le \kappa\}| + |\{Q_t\cap W: Q_t\in\mathcal{F}_\text{right} \text{ and } 1\le |Q_t\cap W|\le \kappa\}| \\
     &\le 2 \cdot C_\text{dom}\cdot|W\cap \pi^{-1}(S_u)|\cdot \kappa^2
    \end{align*}
    (when $u$ is a leaf, the bound is at most $C_\text{dom}\cdot|W\cap \pi^{-1}(S_u)|\cdot \kappa^2$, which trivially satisfies the same inequality).

    Every query sent to the online sampler is processed at some node of the range tree, i.e., the family of queries is $\mathcal{F}=\bigcup_{u}\mathcal{F}_u$. So, summing the traces generated across all nodes $u$ of the range tree, we get
    \begin{align*}
     |\{Q_t &\cap W: Q_t\in\mathcal{F} \text{ and } 1\le |Q_t\cap W|\le \kappa\}| \\
     &\le \sum_{u} |\{Q_t\cap W: Q_t\in\mathcal{F}_u \text{ and } 1\le |Q_t\cap W|\le \kappa\}| \\
     &\le 2C_\text{dom}\kappa^2 \cdot \sum_{u} |W\cap \pi^{-1}(S_u)| \\
     &\le 2C_\text{dom}\kappa^2 \cdot |W| \cdot \log n.
    \end{align*}
    Where the last inequality follows, since every point can lie in at most $O(\log n)$ intervals corresponding to nodes of the range tree.
\end{proof}

We now show that the query sets are heavy.
\begin{lemma}[Heaviness]
\label{lem:heaviness}
For any request $R_t$, the query set $Q_t$ produced by \Cref{alg:bottomless} satisfies $|Q_t| \ge B$.
Furthermore, $Q_t \subseteq A_t(R_t)$, so for any $e \in Q_t$, the point $\pi(e) \in P \cap R_t$, and purchasing $\pi(e)$ hits the request $R_t$.
\end{lemma}
\begin{proof}
    By \Cref{lem:copies}, we know that $|A_t(R_t)| \ge n$. 
    If the node $v$ is a leaf, $Q_t = A_t(R_t)$, and thus $|Q_t| \ge n \ge B$. 
    Otherwise, $v$ is an internal node, and its active copies $A_t(R_t)$ are partitioned into $A_t^{\mathrm{left}}$ and $A_t^{\mathrm{right}}$. Since $Q_t = \arg\max\{|A_t^{\mathrm{left}}|, |A_t^{\mathrm{right}}|\}$, we have $|Q_t| \ge \lceil |A_t(R_t)|/2 \rceil \ge \lceil n/2 \rceil = B$.
    
    In either case, $Q_t \subseteq A_t(R_t)$, so any element $e \in Q_t$ is an active copy of some point in $P \cap R_t$. Thus, purchasing $\pi(e)$ hits $R_t$.
\end{proof}

We now have the ingredients to complete the proof of \Cref{thm:bottomless-rounding}.

\begin{proof}[Proof of \Cref{thm:bottomless-rounding}]
We prove the theorem by showing that \Cref{alg:bottomless} maintains a hitting set $H_t$ with the property that $\Pr[p \in H_t] \leq O(\log \log n) x^t_p$ for every point $p$ and time step $t$. Consider the requests $Q_t$ sent to the online sampler (\Cref{lem:sampler}). By \Cref{lem:heaviness}, we have that $|Q_t| \geq \lceil n/2\rceil$. 
By \Cref{lem:shallow-traces}, the query family satisfies the shallow trace condition with $A = 2C_{\mathrm{dom}}(\lceil \log_2 n \rceil + 1)$. Since $N = n^2$, we have $A \le C_A \log N$ by choosing the constant $C_A = 2C_{\mathrm{dom}} / \ln 2$. 
Together, this allows us to apply \Cref{lem:sampler} with parameters $A = O(\log n)$, $B = \lceil n/2\rceil$, and $N = n^2$ to get that for every point $p \in P$, each of its active copies is in the sampled hitting set $J_t$ with probability at most $O(\frac{1}{n}\log \log n)$. Since $p \in H_t$ only if one of its active copies is in $J_t$ and $p$ has at most $a^t_p \leq 4n x^t_p$ active copies (\Cref{eq:active-copies}), we have that $\Pr[p\in H_t] \leq O(\log \log n) x^t_p$.
\end{proof}

\subsection{Online rounding for unions}
\label{sec:union-rounding}

\begin{definition}
    [Online $\fami$-hitting set]
    Let $U$ be a universe of $n$ elements and $\fami$ be a family of subsets of $U$. An  instance of the \emph{online $\fami$-hitting set problem} is a sequence of requests $R_t \in \fami$. 
\end{definition}

\begin{definition}[$\alpha(n)$-rounding algorithm]
Let $U$ be a universe of $n$ elements and $\fami$ be a family of subsets of $U$. An \emph{$\alpha(n)$-rounding algorithm} receives at time step $t$, the request $R_t \in \fami$ and the current fractional hitting set $x^t$, and maintains an online hitting set $H$ such that $\Pr[e \in H] \leq \alpha(n) \cdot x^t_e$ for every time step $t$ and every element $e$.
\end{definition}

We now show that given rounding algorithms $\ralg_1$ and $\ralg_2$ for $\fami_1$ and $\fami_2$, respectively, we can combine them to get a rounding algorithm for $\fami_3 = \{S_1 \cup S_2 : S_1 \in \fami_1, S_2 \in \fami_2\}$. 

\begin{theorem} \label{thm:union-rounding}
    Let $U$ be a universe of $n$ elements. Suppose $\fami_1$ and $\fami_2$ are two families of subsets of $U$, and let $\fami_3 = \{S_1 \cup S_2 : S_1 \in \fami_1, S_2 \in \fami_2\}$. Furthermore, suppose that $\ralg_1$ is an $\alpha_1(n)$-rounding algorithm for $\fami_1$ and $\ralg_2$ is an $\alpha_2(n)$-rounding algorithm for $\fami_2$. Then, there is a $2(\alpha_1(n) + \alpha_2(n))$-rounding algorithm for $\fami_3$.
\end{theorem}
\begin{proof}
The idea for our algorithm is relatively straightforward. Consider an online $\fami_3$-hitting set instance and let $x$ be an online fractional hitting set. The algorithm runs $\ralg_1$ and $\ralg_2$; let $H_1$ and $H_2$ be their respective hitting sets. At each time step $t$, the algorithm receives a request $R_t \cup S_t$ where $R_t \in \fami_1$ and $S_t \in \fami_2$, and $x^t$, and defines $y^t_e = 2x^t_e$.
If $y^t(R_t) \geq 1$, the algorithm feeds $R_t$ and $y^t$ to $\ralg_1$ and otherwise it feeds $S_t$ and $y^t$ to $\ralg_2$. Then it updates its hitting set $H$ by setting $H = H_1 \cup H_2$.

We now turn to analyzing the above algorithm.
For each time step $t$, since $x$ is a fractional hitting set, we have $x^t(R_t \cup S_t) \geq 1$ 
and so, we get $y^t(R_t) \geq 1$ or $y^t(S_t) \geq 1$. Thus, $y^t$ is indeed a fractional hitting set for the sequence of requests sent to $\ralg_1$ and also for the sequence of requests sent to $\ralg_2$.  
The guarantees of $\ralg_1$ and $\ralg_2$ imply that for each element $e \in U$ and each time step $t$, we have $\Pr[e \in H_1] \leq \alpha_1(n) \cdot y^t_e$ and $\Pr[e \in H_2] \leq \alpha_2(n) \cdot y^t_e$. Since $H = H_1 \cup H_2$, we have that $\Pr[e \in H] \leq (\alpha_1(n)+\alpha_2(n))\cdot y^t_e \le 2(\alpha_1(n)+\alpha_2(n))\cdot x^t_e$, as desired.
\end{proof}

\begin{proof}[Proof of \Cref{thm:TL-rounding}]
    By rotation of the coordinate axes, \Cref{thm:bottomless-rounding} holds identically for topless and leftless rectangles, yielding an $O(\log \log n)$-rounding algorithm for each family. By \Cref{thm:union-rounding}, these can be combined into an $O(\log \log n)$-rounding algorithm for TL-rectangles (which are unions of topless and leftless rectangles).
    
    Given a $\rho(n)$-competitive fractional solution $x$, this rounding algorithm maintains a feasible integral hitting set $H$ such that $\Pr[p \in H] \le O(\log \log n) x_p$ for all $p \in P$. By linearity of expectation, the expected cost of $H$ satisfies:
    \[
        \mathbb{E}[c(H)] = \sum_{p \in P} c_p \Pr[p \in H] \le O(\log \log n) \sum_{p \in P} c_p x_p \le O(\rho(n) \log \log n) \cdot \mathrm{OPT}. \qedhere
    \]
\end{proof}

\section{Unweighted OCycAP}\label{sec:unweighted-ocycap}

In this section, we prove \Cref{thm:NOCAP} by presenting an algorithm achieving a $10$-competitive ratio for the OCycAP problem with node arrivals in \Cref{sec:NOCAP}, and a $\Omega(\log n)$ lower bound in the cut-arrivals setting in \Cref{sec:ocycap-cut-lb}.

\subsection[O(1)-competitive for node arrivals]{$O(1)$-competitive for node arrivals}
\label{sec:NOCAP}

We now describe a deterministic $O(1)$-competitive algorithm for OCycAP with node arrivals. 

\begin{theorem}\label{thm:constant-node-arr}
    There is a deterministic 10-competitive algorithm for OCycAP with node arrivals.
\end{theorem}

\subsubsection{Notation and Terminology}

To describe the algorithm, we first establish some notation. Since the given graph $R$ is a cycle, we call two links $\ell$ and $\ell'$ \emph{intersecting} if they intersect when drawn as straight lines on a planar circular embedding of $R$ (including if they intersect at an endpoint). The \emph{link intersection graph} of $R$ is then defined as the graph on a node set corresponding to the links $L$, and with two links adjacent whenever they are intersecting. Recall that in the node arrival setting, our goal is to purchase links so that every requested terminal pair $(s, t)$ is satisfied: it becomes 3-edge-connected in the augmented graph. It is well-known (see, e.g., \cite{BasavarajuEtAl2014, TraubZenklusen2023WCAP}) that a pair is satisfied if and only if there is a path of purchased links in the link intersection graph from a link with $s$ as an endpoint to a link with $t$ as an endpoint. 

Therefore, it useful to study connected component of the link intersection graph for the set of links chosen by the algorithm. For a component $S$, we abuse notation and say it contains a vertex $v$ of $R$ if any link in $S$ is incident to $v$. 

For any vertex $v$ on the cycle, consider the clockwise labeling $v=v_0, \ldots, v_{n-1}$ of the vertices of the cycle $R$. Given a set $S$ of links, define its \emph{interval with respect to $v$}, denoted $I_v(S)$, to consist of the vertices $v_i, v_{i+1}, \ldots, v_j$ where $i,j$ are the smallest and largest indices of the vertices incident to any link in $S$. Note that even if $S$ is a connected component in the link intersection graph, $I_v(S)$ may contain vertices that are not in the same connected component as $S$.\footnote{For example, suppose $S$ consists of a single link $(v_0,v_{n-1})$, then $I_v(S)$ contains every vertex.} The following lemma establishes additional structure on the intervals of connected components of the link intersection graph. 

\begin{lemma}\label{lem:laminar}
For any link intersection graph corresponding to a set of links on $R$, the set of intervals of connected components forms a laminar family. 
\end{lemma}
\begin{proof}
    Observe that if $I_v(S) = [i,j]$ then for every $i'$ such that $i \leq i' < j$, there is a link  $\ell \in S$ such that $I_v(\ell)$ contains $(i',i'+1)$. Thus, if the intervals $I_v(S)$ and $I_v(S')$ cross, then there is a link $\ell \in S$ and a link $\ell' \in S'$ whose intervals $I_v(\ell)$ and $I_v(\ell')$ cross. Thus, $\ell$ and $\ell'$ are adjacent to each other in the link intersection graph.
\end{proof}

\subsubsection{Algorithm for Node Arrivals}

With this terminology in hand, we can now define the algorithm. In \Cref{alg:node-arrival-alg}, we say that a connected component of the link intersection graph is \emph{active} if it contains any terminal seen so far. 

\SetKwBlock{OnlineArrival}{When terminal pair $(s,t)$ arrives:}{} 
\begin{algorithm}[ht]	
    \caption{\texttt{OCycAP with Node Arrival}}
    \label{alg:node-arrival-alg}
	\Input{Cycle $R$, collection $L$ of links, and a sequence $(s_i, t_i) \in \binom{V}{2}$ of terminal pairs.}
	\Output{A set of links so that so that $(s_i, t_i)$ are 3-edge-connected.}
    \BlankLine

    \OnlineArrival{
        Let $S$ and $T$ be the active components containing $s$ and $t$, respectively.\;
        Suppose WLOG (by \Cref{lem:laminar}, swapping the labels of $s$ and $t$ if necessary) that either $I_t(S)$ and $I_t(T)$ are disjoint or $I_t(S)$ is contained in $I_t(T)$.\;
        \While{$S \neq T$}{
        (Merge step) If there exist at most 2 links such that buying them merges $S$ with another active component, buy them.\;
        (Bracketing step) Otherwise, buy the two links leaving $I_t(S)$ whose endpoints outside of $I_t(S)$ have the smallest and largest index (with respect to the ordering with $v_0 = t$).\;
        }
    }
\end{algorithm}

To prove \Cref{thm:constant-node-arr}, we show that the total cost of merge steps is at most $8\OPT$ and the total cost of bracketing steps is at most $2\OPT$ for a total cost of at most $10\OPT$.

\begin{proof}[Proof of \Cref{thm:constant-node-arr}]
Let $r$ be the total number of distinct terminals requested across all arrivals. Since every valid solution must cover all $r$ distinct terminals, and each link is incident to at most two endpoints, we have that $\nicefrac{r}{2} \leq \OPT$. Moreover, each merge step strictly reduces the number of active components by one, so starting from at most $r$ initial active components, there can be at most $r-1$ merge steps. Thus, the total cost due to merge steps is at most $2(r-1) \leq 4\OPT \leq 8\OPT$.

We now define a charging scheme to show that the number of bracketing steps is at most $\OPT$.
Consider an iteration of the algorithm in which a bracketing step was performed on interval $I_t(S)$. Let $A$ be the set of links bought by the algorithm so far, and let $P^*$ be the shortest path in $\OPT$ that connects the component $S$ and $t$. Since the merge step was not taken in this iteration of the algorithm, we have that $|P^*| \geq 3$. We will charge the bracketing step to a link of $P^*$ that is newly incident to $A$.

\begin{claim}
    After each bracketing step, there is a link of $\OPT$ that is newly incident to $A$.
\end{claim}

\begin{proof}
    Let $S'$ be the component containing $S$ after the bracketing step. Since the bracketing step buys the two links $\ell_1,\ell_2$ leaving $I_t(S)$ whose endpoints have the smallest and largest index with respect to $v_0 = t$, respectively, it follows that $I_t(S')$ contains both endpoints of all links incident to $I_t(S)$. Hence, the first link of $P^*$ has an endpoint in $I_t(S)$ and another endpoint in $I_t(S') \setminus I_t(S)$. Moreover, as $P^*$ is the shortest path between $S$ and $t$, only the first link of $P^*$ has an endpoint in $I_t(S)$. Finally, observe that $I_t(S')$ cannot contain $t$, since $v_0 = t$ would then imply that one of the links chosen is incident to $t$, and so the algorithm would have instead taken a merge step. Thus, there exists a link $\ell^* \in P^*$ with one endpoint in $I_t(S') \setminus I_t(S)$ and the other outside of $I_t(S')$.

    Observe that $\ell^*$ is not incident to $S$. It is incident to at least one of $\ell_1$ and $\ell_2$, and so it cannot be incident to any other active component since otherwise, the algorithm would have taken a merge step instead of a bracketing step. Therefore, $\ell^*$ is newly incident to $A$.
\end{proof}

Overall, there are at most $\OPT$ many bracketing steps, each costing two links, so the total cost of the algorithm is at most $8\OPT + 2\OPT = 10\OPT$. 
\end{proof}

\subsection[An Omega(log n) lower bound for OCycAP with cut-arrivals]{An $\Omega(\log n)$ lower bound for OCycAP with cut-arrivals}
\label{sec:ocycap-cut-lb}
As discussed previously in \Cref{sec:techniques}, there is no $o(\log n)$-competitive algorithm for OCycAP with cut-arrivals, as it captures general OCAP. We also give a simple direct proof that also rules out $o(\log n)$-competitive fractional algorithms below.

\begin{theorem}
There is no fractional algorithm for OCycAP with cut arrivals that achieves a competitive ratio better than $\log n$. 
\end{theorem}

\begin{proof}
Let $R = (V,E)$ be a cycle with $n$ nodes $v_0, \ldots, v_{n-1}$. Let $L = \binom{V}{2}$. Recall from \Cref{sec:ocycap} that a min cut $C$ of $R$ can be specified as an interval $[a,b]$ where $1 \leq a \leq b \leq n-1$ where $\{v_a, \ldots, v_b\}$ is one of the two sides of the cut.

Fix a fractional algorithm and let $x^t_\ell$ be the extent to which it purchased link $\ell$ in response to the first $t$ cuts. WLOG, we assume that it is lazy, i.e., $x^t_\ell > 0$ only if $\ell$ covers a cut $C_{t'}$ for $t' \leq t$. The first cut is $C_1 = [1,n/2]$. Each subsequent cut $C_t$ for $t > 1$ is obtained as follows. Let $C_{t-1} = [a_{t-1}, b_{t-1}]$ be the previous cut and $m_{t-1} = \lfloor \frac{a_{t-1} + b_{t-1}}{2} \rfloor$ be the median of $[a_{t-1}, b_{t-1}]$. Let $A_{t-1} = [a_{t-1}, m_{t-1}]$ and $B_{t-1} = [m_{t-1} + 1, b_{t-1}]$. If $\sum_{\ell:\,\ell\text{ covers }A_{t-1}} x^{t-1}_\ell \leq \sum_{\ell:\,\ell\text{ covers }B_{t-1}} x^{t-1}_\ell$ then $C_t = A_{t-1}$ otherwise $C_t = B_{t-1}$. There are a total of $r = \log_2 n - 1$ requests that form a nested family of intervals with the last request being a singleton $[a_r]$.

Observe that $\OPT = 1$ as the link $(v_0, v_{a_r})$ satisfies all cuts. On the other hand, since $x$ is lazy, we have that 
\[\sum_{\ell:\,\ell\text{ covers }A_{t-1}} x^{t-1}_\ell
+\sum_{\ell:\,\ell\text{ covers }B_{t-1}} x^{t-1}_\ell
=\sum_{\ell:\,\ell\text{ covers }C_{t-1}} x^{t-1}_\ell \geq 1\] so by our choice of $C_t$ \[\sum_{\ell:\,\ell\text{ covers }C_{t-1}} x^{t-1}_\ell - \sum_{\ell:\,\ell\text{ covers }C_{t}} x^{t-1}_\ell \geq 1/2.\]
Summing over $t$ gives us that the cost of the fractional solution is $\Omega(\log n)$.
\end{proof}

\section*{Independent Work}
During a workshop in August 2026, we discovered that another group consisting of Garg, Hommelsheim, and Lindermayr were working on Online Connectivity Augmentation, and that they had independently obtained an $O(1)$-competitive algorithm for OCycAP with node-arrivals. No techniques or other results were discussed, and we continued working independently following this point.

\section*{AI Disclosure}
The authors used Gemini 3.8 Flash and Gemini 3.1 Pro-preview for minimal proofreading and editing. The authors also used ChatGPT 5.6 - 6.1 in the development of some of the algorithmic ideas; in particular, the connection to geometric hitting set and the constant-competitive algorithm for OCycAP with node-arrivals was discovered by ChatGPT 6, which also assisted us in adapting the methods of \cite{bhore2026} to our setting. ChatGPT 6 was also used to create tikz diagrams, for example in \Cref{fig:cycle-geometric-reduction}. The authors verified the correctness and originality of all content, including references.

\AtNextBibliography{\small}
\printbibliography

\appendix

\section{Proof of \Cref{thm:octant-traces}}
\label{app:octant-traces}

Let $W \subset \mathbb{R}^3$ be a set of $r$ points, and let $\mathcal{G}$ denote the family of octants of a fixed orientation. Without loss of generality, assume these are standard 3D dominance ranges of the form $Q_q = (-\infty, q_x] \times (-\infty, q_y] \times (-\infty, q_z]$ defined by a query vertex $q = (q_x, q_y, q_z)$. We wish to bound the number of distinct shallow traces $Q_q \cap W$ of size at most $\kappa$.

To do this, we rely on the existence of optimal shallow cuttings for 3D dominance ranges, as established by Afshani and Tsakalidis \cite{AfshaniT14}. A \emph{$\kappa$-shallow cutting} for $W$ is a collection of spatial cells (regions) $\mathcal{C}$ with the following properties:
\begin{enumerate}
    \item \textbf{Covering:} The union of the cells in $\mathcal{C}$ completely covers the $\kappa$-shallow region of $\mathbb{R}^3$. That is, for any query $q$ such that $|Q_q \cap W| \le \kappa$, the vertex $q$ lies in at least one cell $C \in \mathcal{C}$.
    \item \textbf{Size:} The number of cells in the cutting is $|\mathcal{C}| = O(r / \kappa)$.
    \item \textbf{Conflict bound:} For each cell $C \in \mathcal{C}$, let $W_C \subseteq W$ be the subset of points that can be dominated by any query $q \in C$. The cutting guarantees that $|W_C| = O(\kappa)$.
\end{enumerate}

Every query octant $Q_q$ that captures at most $\kappa$ points has its vertex $q$ located inside some cell $C \in \mathcal{C}$. Consequently, the trace of $Q_q$ on $W$ is entirely a subset of $W_C$, meaning $Q_q \cap W = Q_q \cap W_C$. Thus, to bound the total number of distinct shallow traces of size at most $\kappa$ over all of $\mathbb{R}^3$, it suffices to bound the number of distinct traces that can be formed within each cell $C$, and sum this over all cells in $\mathcal{C}$.

For a fixed cell $C$, the number of points interacting with queries in $C$ is $|W_C| = O(\kappa)$. Any distinct trace generated by an octant $Q_q$ can be uniquely identified by its minimal bounding octant. Because an octant is defined by exactly 3 orthogonal planes, this minimal bounding octant is pinned down by at most 3 extreme points from the trace (the points with the maximum $x$, $y$, and $z$ coordinates, respectively). 

Therefore, the number of distinct traces that can be formed on $W_C$ is bounded by the number of subsets of $W_C$ of size at most 3. This gives:
\[
    \text{Number of traces in } C \le \sum_{i=1}^{3} \binom{|W_C|}{i} = O(|W_C|^3) = O(\kappa^3).
\]

Summing this bound over all $|\mathcal{C}| = O(r / \kappa)$ cells in the shallow cutting, the total number of distinct traces of size at most $\kappa$ is at most:
\[
    \sum_{C \in \mathcal{C}} O(\kappa^3) = O\left(\frac{r}{\kappa}\right) \cdot O(\kappa^3) = O(r \kappa^2).
\]

Letting $C_{\mathrm{dom}}$ be the absolute constant absorbing the asymptotic notation, we conclude that the number of distinct nonempty traces of size at most $\kappa$ is at most $C_{\mathrm{dom}} r \kappa^2$, completing the proof.
\qed

\end{document}